\documentclass[letterpaper, 10 pt, conference, letterpaper]{ieeeconf}  

\IEEEoverridecommandlockouts                              

\usepackage{amsmath,amssymb,amsfonts,mathrsfs,indentfirst,graphicx,float}

\newtheorem{assumption}{Assumption}

\allowdisplaybreaks[4]

\title{\LARGE \bf
Koopman-type inverse operator for linear non-minimum phase systems with disturbances*
}

\author{Yuhan Li, and Xiaoqiang Ji*, $\textit{IEEE Member}$
\thanks{* This work was partially supported by Shenzhen Science and Technology Program (Grant No. RCBS20210706092219050), Guangdong Basic and Applied Basic Research Foundation (Grant No. 2022A1515110411, Grant No. 2023A1515012883), and Shenzhen Institute of Artiﬁcial Intel-ligence and Robotics for Society (AC01202201001).}
\thanks{Y.Li, and X. Ji are with the Shenzhen Institute of Artificial Intelligence and Robotics for Society, Shenzhen 518172, China, and the School of Science and Engineering, The Chinese University of Hong Kong, Shenzhen, Shenzhen 518172, China.}
\thanks{* Corresponding author, email: jixiaoqiang@cuhk.edu.cn}%
}

\begin{document}

\maketitle
\thispagestyle{empty}
\pagestyle{empty}
\UseRawInputEncoding

\begin{abstract}

In this paper, a novel Koopman-type inverse operator for linear time-invariant non-minimum phase systems with stochastic disturbances is proposed. This operator employs functions of the desired output to directly calculate the input. Furthermore, it can be applied as a data-driven approach for systems with unknown parameters yet a known relative degree, which is a departure from the majority of existing data-driven methods that are only applicable to minimum phase systems. Based on this foundation, we use the Monte Carlo approach to develop an improved Koopman-type method for addressing the issue of inaccurate parameter estimation in data-driven systems with large disturbances. The simulation results justify the tracking accuracy of Koopman-type operator.
\end{abstract}

\section{INTRODUCTION}

Numerous studies have demonstrated that output tracking for non-minimum phase (NMP) systems poses more significant challenges than minimum phase (MP) systems. A system is considered to have non-minimum phase (or possess unstable zeros in the linear case) if there exists a (nonlinear) state feedback capable of maintaining the system output at an identical zero level, while simultaneously causing the internal dynamics to become unstable (Isidori and Alberto, 1985)$\nocite{isidori1985nonlinear}$. For linear systems, the inverse of the system will make the unstable zeros of the original system transfer function become the poles, which cause the instability of the inverse (Butterworth et al., 2018)$\nocite{butterworth2008effect}$.
$\nocite{devasia1996nonlinear}$$\nocite{devasia1998stable}$$\nocite{estrada2021toward}$

Various approaches have been proposed to tackle the difficulties in NPM systems, including decomposing the system into external and internal dynamics and solving for the internal dynamics (Devasia et al., 1996; Devasia and PadenFrom, 1998; Estrada, 2021)$\nocite{devasia1996nonlinear}$$\nocite{devasia1998stable}$$\nocite{estrada2021toward}$. Berget and Tomas (2020)$\nocite{berger2020tracking}$ proposed a method for constructing a new output that eliminates the unstable part of the zero dynamics, while Zundert et al. (2019)$\nocite{van2019stable}$ have decomposed NMP systems into a stable part and an unstable part for separate control. Ma et al. (2020)$\nocite{ma2020dual}$ solved the problem of random perturbations in NMP systems by using two cost functions that possess the dual property. However, one of the main disadvantages of model-based control methods is their reliance on accurate models of the systems, which can be difficult and time-consuming to develop and validate for complex NMP systems.

In recent years, data-driven approaches have garnered attention for their ability to address the challenge of obtaining accurate models for model-based control strategies. Among the classic methods for non-minimum phase systems are virtual reference feedback tuning (VRFT) (Campi et al., 2002) and correlation-based tuning (CBT) (Van Heusden et al., 2011). However, CBT and VRFT require a comprehensive dataset to ensure model accuracy and may rely on the selection of certain empirical parameters, as noted by Rallo et al. (2016). To overcome these limitations, Suresh Kumar et al. (2022) developed a data-driven control method that integrates internal model control and VRFT and designed a generalizable methodology for data-driven identification of nonlinear dynamics based on the Koopman operator. Markovsky et al. (2022) proposed and solved the data-driven dynamic interpolation and approximation problem. However, most data-driven NMP control methods often require large amounts of data for training, and their applicability conditions and control accuracy have not been mathematically proven. Mamakoukas et al. (2021) designed a generalizable methodology for data-driven identification of nonlinear dynamics based on the Koopman operator.

In this paper, we propose a data-driven method that requires very few parameters and training data and provide a proof for the tracking accuracy. For systems without disturbances, only a small period of input-output data is needed to determine the parameters of our method.

The Koopman method, initially introduced by Koopman (1931), has proven to be a valuable mathematical tool in transforming complex nonlinear systems into higher-dimensional linear systems. This transformation enables the application of established linear system theory to control nonlinear systems. In recent years, the Koopman method has gained significant attention in various fields, including cybernetics and machine learning. Mauroy et al. (2021) introduced the use of Koopman operators in control systems. Mamakoukas et al. (2020) demonstrated the effectiveness of the Koopman operator for a robotic system at the experimental level. Additionally, Klus et al. (2020) presented an application of the Koopman operator in data-driven methods, and Leon and Devasia (2022) proposed a Koopman-type data-driven approach to control linear MP systems. Further research on the Koopman method holds the potential to revolutionize the field of nonlinear control systems.
$\nocite{ampi2002virtual}$$\nocite{van2011data}$$\nocite{rallo2016data}$$\nocite{sureshrobust}$$\nocite{mamakoukas2021derivative}$$\nocite{markovsky2022data}$
$\nocite{koopman1931hamiltonian}$$\nocite{mauroy2020introduction}$$\nocite{mamakoukas2021derivative}$$\nocite{klus2020data}$$\nocite{yan2022precision}$

The Koopman method has been widely studied in control systems, with research falling into two main categories: selecting eigenfunctions of the Koopman operator experimentally and obtaining desired coefficients using machine learning methods, or introducing a Koopman method under an MP system and proving its applicability and effectiveness. However, there is currently no proof of the practicality of the Koopman method for NMP systems. In this paper, we propose a Koopman-type control method for linear NMP systems with stochastic disturbances and prove its applicability conditions and tracking effects. The contributions of this paper are twofold: first, to the best of our knowledge, this paper is the first to use the Koopman method for the control of linear NMP systems and provide a complete proof; second, we demonstrate that our Koopman-type operator is still applicable for systems with random perturbations.

The remainder of the paper is organized as follows, the introduction of basic notation and the Koopman method is in Section II, the problem of NMP system control is in Section III, the Koopman-type method is proposed in Section IV with the theoretical analysis and the demonstration of tracking error. Section V includes simulation results justifies the tracking accuracy theorem presented in Chapter IV. A brief conclusion is in Section IV.

\section{PRELIMINARIES}
\noindent
$\\ \textbf{A. Notation}$

Throughout this paper, let $\mathbb{R}$ denotes the set of real numbers,  $\mathbb{R}^{m\times n}$ denotes the set of $m\times n$ matrix, and $\mathbb{Z}^+$ denotes the set of positive integers. For vector $\mathbf{a}$, $\mathbf{b}$, $||\mathbf{a}||$ is the $l_2$ norm, dim($\mathbf{a}$) is the dimension of vector $\mathbf{a}$, $<\mathbf{a},\mathbf{b}>$ is the inner product, . For matrix $\mathbf{A}$, $||\mathbf{A}||$ is the matrix Frobenius norm, $\mathbf{A}^T$ is the transpose of matrix $\mathbf{A}$, $\mathbf{A}^\dagger$ is the Moore-Penrose pseudoinverse of matrix $\mathbf{A}$. $U[x]$ is a uniform distribution on interval $[-x,x]$ for $x\in\mathbb{R}^+$. For any matrix $\mathbf{A}=(a_{ij})$, where $a_{ij}$ are random variables, then $E[\mathbf{A}]=(E[a_{ij}])$ is the mathematical expectation of matrix $\mathbf{A}$. $y^{(r)}(t)$ is the r-th derivative of $y$ respect to $t$.$\\$
\noindent
$\\ \textbf{B. Koopman method}$

	Consider a (Banach) space $\mathscr{F}$ of observables $f:\boldmath{X}\rightarrow\mathbb{C}$. The Koopman operator $U_s:\mathscr{F}\rightarrow\mathscr{F}$ associated with the map $\boldmath{S}:\boldmath{X}\rightarrow \boldmath{X}$ is defined through the composition (Mauroy et al., 2020)$\nocite{mauroy2020introduction}$
	\begin{equation}
		U_sf=f\circ \boldmath{S}\ \quad f\in \mathscr{F}
	\end{equation}
$U_s$ is Koopman operator which convert nonlinear systems to finite or infinite dimensional linear systems by decomposition of eigenequations and eigenvalues of $U_s$ and treat the nonlinear system as a higher order linear system.

The conventional Koopman method necessitates the identification of specific observables, such that they constitute a linear system. Subsequently, the linear system composed of these observables is analyzed to derive the control strategy for the linear system. The final step involves solving the inverse of the observables to obtain the desired parameters of the original system, such as the input function $u(t)$ of the system. A noteworthy concept inspired by the Koopman approach is the potential to directly and linearly acquire the desired parameters through observables.  A valuable idea inspired by Yan and Devasia (2022)$\nocite{yan2022precision}$ is along the lines of the Koopman approach is the possibility of getting the parameters we want directly and linearly through observables. This modified Koopman method we call Koopman-type method.

\section{PROBLEM STATEMENT}

Consider a non-minimum phase (NMP) stochastic linear time-invariant (LTI) system defined by
\begin{equation}
	\begin{aligned}
		\dot{\mathbf{x}}(t)&=\mathbf{A}\mathbf{x}(t)+\mathbf{B}u(t)+\mathbf{G}w(t)\\
		y(t)&=\mathbf{C}\mathbf{x}(t)+h(t)
	\end{aligned}		
\end{equation}
Where states $\mathbf{x}(t)\in \mathbb{R}^n$, outpute $y(t)\in \mathbb{R}$, $h(t)$ and $w(t)$ are stochastic disturbance functions, $\mathbf{A}\in \mathbb{R}^{n\times n}$, $\mathbf{B},\mathbf{G}\in\mathbb{R}^{n\times 1}$, and $\mathbf{C}\in \mathbb{R}^{1\times n}$. 
$\\ \textbf{Assumption 1: }$
	$h(t)$, $w(t)$ are bounded, i.e. $\sup\limits_{t\in \mathbb{R}}|h(t)|<\infty$ and $\sup\limits_{t\in \mathbb{R}}|w(t)|<\infty$

Perform Laplace transform on (1),
\begin{equation}
	\begin{aligned}
		Y(s)=&[\mathbf{C}(s\mathbf{I}_n-\mathbf{A})^{-1}\mathbf{B}]U(s)+\\&[\mathbf{C}(s\mathbf{I}_n-\mathbf{A})^{-1}\mathbf{G}]W(s)+H(s)
	\end{aligned}
\end{equation}
Write the first part of (3) in the form
\begin{equation}
	\begin{aligned}
		&\mathbf{C}(s\mathbf{I}_n-\mathbf{A})^{-1}\mathbf{B}=\\&k\cdot \frac{s^{n-r}+b_{n-r-1}\cdot s^{n-r-1}+...+b_0}{s^n+a_{n-1}s^{n-1}+...+a_1s+a_0}
	\end{aligned}
\end{equation}
$\textbf{Assumption 2: }$
	The relative degree $r\leq n$ is known and the exact value of matrix $\mathbf{A},\mathbf{B},\mathbf{C},\mathbf{G}$ are unknown.
$\\ \textbf{Assumption 3: }$
	The polynomial in the numerator of (4) has roots with real parts greater than zero but no roots with real parts equal to zero, while the roots of the denominator polynomial have all real parts less than zero.

Take $y_e=y(t)-h(t)$, $\mathbf{\xi}=(y_e,\dot{y}_e,\ddot{y}_e,...,y_e^{(r-1)})^T$, $\mathbf{\eta}=(x_1,x_2,...,x_{n-r})^T$, we call $\eta$ the internal states, where $x_i$ is the i-th element in $\mathbf{x}$. Using linear transformations the same as in Hendricks et al. (2008)$\nocite{hendricks2008linear}$, the system can be written as
\begin{equation}
	\begin{aligned}
		\dot{\mathbf{\xi}}(t)&=\mathbf{A}_1\mathbf{\xi}(t)+\mathbf{A}_2\mathbf{\eta}(t)+\mathbf{B}_1u(t)+\mathbf{G}_1w(t)\\
		\dot{\mathbf{\eta}}(t)&=\mathbf{A}_3y(t)+\mathbf{A}_4\mathbf{\eta}(t)+\mathbf{G}_2w(t)
	\end{aligned}
\end{equation}
Where
\begin{equation}
	\begin{aligned}
		\mathbf{A}_3=(0,0,...,1)^T\in\mathbb{R}^{n\times 1}\\
		\mathbf{B}_1=(0,0,...,k)^T\in\mathbb{R}^{n\times 1}\\
		\mathbf{G}_1=(0,0,...,g)^T\in\mathbb{R}^{n\times 1}
	\end{aligned}
\end{equation}
and 
\begin{equation}
	\begin{aligned}
		\mathbf{A}_1&=
		\left ( \begin{array}{ccc}
			0 &\cdots &0\\
			\vdots &\vdots &\vdots\\
			- &\mathbf{r}^T &-\\
		\end{array} \right )
		\mathbf{A}_2=
		\left ( \begin{array}{ccc}
			0 &\cdots &0\\
			\vdots &\vdots &\vdots\\
			- &\mathbf{s}^T &-\\
		\end{array} \right )\\
		\mathbf{A}_4&=
		\left ( \begin{array}{cccc}
			0 &1 &0 &\cdots\\
			\vdots &\ddots &\ddots &\vdots\\
			0 &\cdots &0 &1\\
			-b_0 &-b_1 &\cdots &-b_{n-r-1}
		\end{array} \right )
	\end{aligned}
\end{equation}
Denote $y_d(t)$ as the desired output, the task of this paper is to find a Koopman-type operator to calculate input $u(t)$ directly from $y_d$ such that under this input, the actual output $y(t)$ and $y_d(t)$ differ by an acceptable error as time grows.
\begin{assumption}
	$y_d$ is r-order continuously differentiable and there is $M>0$, $y_d(t)\leq M$ for any $t\in \mathbb{R}$.
\end{assumption}

\section{MAIN RESULTS}

In this section, we describe the Koopman-type operator in detail. In part A, we present the specific formulation of the Koopman-type method and conduct a theoretical analysis of its tracking accuracy. In part B, we introduce a data-driven approach for determining the parameters required by the Koopman-type method. In part C, we designed an improved Koopman-type method by incorporating the Monte Carlo technique to improve the accuracy of parameters estimation in data-driven processes for systems with large disturbances.$\\ \\$
$\textbf{A. Koopman-type operator and its performance analysis}\\ \\$
$\textbf{Definition 1}$ Koopman-type operator
\begin{equation}
	\begin{aligned}
		\mathcal{A}:=&\{q^{-d\cdot\Delta t}y_d(t)|d\in \{0,\pm1\cdots\pm N-1,N\}\}\\\mathcal{D}:=&\{y_d^{(i)}|i\in\{1,2\cdots r\}\}
	\end{aligned}
\end{equation}
Where $q^{-d\cdot\Delta t}$ is $d\cdot\Delta t$ pure time delay. Take $\Phi$ be a column vector containing every function in $\mathcal{A}$ and $\mathcal{D}$, we define the Koopman-type operator in the form
\begin{equation}
	u(t)=<\mathbf{K},\Phi(t)>
\end{equation}
Where $\mathbf{K}$ is a constant column vector with dimension $2N+r$.
	$\\\textit{Remark 1:}$
	$\mathbf{K}$ is an undetermined parameter of the Koopman-type operator, for different systems we need to determine different $\mathbf{K}$ to improve the accuracy of tracking.
	$\\ \\ \textbf{Theorem 1: }$
	There exists $\delta>0$, for any $\epsilon>0$, there exist $\hat{N}$, $\Delta \hat{t}$ and $\mathbf{K}$, if $N>\hat{N}$, and $\Delta t<\Delta \hat{t}$, then from the Koopman-type operator, the output of system (1) satisfies
	\begin{equation}
		|y(t)-y_d(t)|\leq\epsilon+\delta\sup\limits_{t\in \mathbb{R}}|w(t)|+\sup\limits_{t\in \mathbb{R}}|r(t)|
	\end{equation}
	for large enough t.

$\\$
$\textbf{Proof}$
By the linear variable substitution of $\mathbf{x}$, Zou and Devasia (1999) $\nocite{zou1999preview}$, the second equation of system (3) can be written as
\begin{equation}
	\dot{\eta'}(t)=\mathbf{A}_{3'}y_e(t)+\mathbf{A}_{4'}\eta'(t)+\mathbf{G}_{2'}w(t)
\end{equation}
Where
\begin{equation}
	\mathbf{A}_{4'}=
	\begin{bmatrix}
		\mathbf{A}_{4'-} & 0\\
		0 & \mathbf{A}_{4'+}
	\end{bmatrix}
\end{equation}
Where all eigenvalues of $\mathbf{A}_{4'-}(\mathbf{A}_{4'+})$, have negative (positive) real part. Without of lose generality, assume $\mathbf{A}_4$ is directly this form. From Devasia (1996)$\nocite{devasia1996nonlinear}$, the unique solution of the second equation in system (3) with the assumption $\eta(\pm\infty)=0$ for given $y_d$ is
\begin{equation}
	\eta(t)=\int_{-\infty}^{+\infty}\phi(t-\tau)(\mathbf{A}_3y_d(\tau)+\mathbf{G}_2w(\tau))d\tau
\end{equation}
Where
\begin{equation}
	\phi(t)=
	\begin{bmatrix}
		1(t)e^{\mathbf{A}_{4-}} & 0\\
		0 & -1(-t)e^{\mathbf{A}_{4+}}
	\end{bmatrix}
\end{equation}
	$\textbf{Lemma 1: }$
	There exists positive scalars $\alpha>0,\beta>0,\gamma>0$ such that
	\begin{equation}
		||\eta(t)-\eta_N(t)||\leq \beta e^{-\alpha N\Delta t}+\gamma\sup\limits_{t\in \mathbb{R}}|w(t)|
	\end{equation}
	where $\eta(t)$ is calcullated in (13), and
	\begin{equation}
		\eta_N(t)=\int_{t-N\Delta t}^{t+N\Delta t}\phi(t-\tau)(\mathbf{A}_3y_d(\tau)+\mathbf{G}_2w(\tau))d\tau
	\end{equation}

$\\$
$\textbf{Proof}$ The eigenvalues of $\mathbf{A}_{4-}$ and $-\mathbf{A}_{4+}$ have negative real parts so there exists postive scalars $\kappa_1> 0$, $\kappa_2>0$, $\kappa_3>0$, $\alpha_1>0$, $\alpha_2>0$, $\alpha_3>0$ such that, see Desoer and Vidyasagar (2009) $\nocite{desoer2009feedback}$
\begin{equation}
	\begin{aligned}
		||\phi(t)||&\leq\kappa_1e^{-\alpha_1t}\\
		||\phi(-t)||&\leq\kappa_2e^{-\alpha_2t}\\
		||e^{\mathbf{A}t}||&\leq\kappa_3e^{-\alpha_3t}
	\end{aligned}
\end{equation}

Then
\begin{align*}
	||\eta(t)-\eta_N(t)||=&||\int_{-\infty}^{t-N\Delta t} \phi(t-\tau)\mathbf{A}_3y_d(\tau)d\tau+\\&\int_{t+N\Delta t}^{\infty} \phi(t-\tau)\mathbf{A}_3y_d(\tau)d\tau+\\&
	\int_{-\infty}^{+\infty}\phi(t-\tau)\mathbf{G}_2w(\tau)d\tau||\\
	\leq & M||\mathbf{A}_3||(||\int_{-\infty}^{t-N\Delta t} \kappa_1e^{-\alpha_1(t-\tau)}d\tau||+\\&||\int_{t+N\Delta t}^{\infty} \kappa_2e^{-\alpha_2(t-\tau)}d\tau||)+\\
	&\sup\limits_{t\in \mathbb{R}}|w(t)|\cdot||\mathbf{G}_2||\\&(||\int_{-\infty}^{t} \kappa_1e^{-\alpha_1(t-\tau)}d\tau||+\\&||\int_{t}^{\infty} \kappa_2e^{-\alpha_2(t-\tau)}d\tau||)\\
	=&M||\mathbf{A}_3||(\frac{\kappa_1}{\alpha_1}e^{-\alpha_1N\Delta t}+\frac{\kappa_2}{\alpha_2}e^{-\alpha_2N\Delta t})+\\
	&\sup\limits_{t\in \mathbb{R}}|w(t)|\cdot||\mathbf{A}_2||(\frac{\kappa_1}{\alpha_1}+\frac{\kappa_2}{\alpha_2})
\end{align*}
Take $\alpha=\min{\alpha_1,\alpha_2}$, $\beta=2M||\mathbf{A}_3||\max{\frac{\kappa_1}{\alpha_1},\frac{\kappa_2}{\alpha_2}}$, $\gamma=||\mathbf{G}_2||(\frac{\kappa_1}{\alpha_1}+\frac{\kappa_2}{\alpha_2})$ $\blacksquare$

By the definition of integration, we can write the integral as partial sum
\begin{equation}
	\begin{aligned}
		\eta_N(t)=&\int_{t-N\Delta t}^{t+N\Delta t} \phi(t-\tau)\mathbf{A}_3y(\tau)d\tau\\
		\approx&\Delta t\cdot\sum_{\tau=0}^{2N-1}\phi((-N+\tau)\Delta t)\mathbf{A}_3q^{(N-\tau-1)\Delta t}y_d(t)\\
		\triangleq&\hat{\eta}_N(t)
	\end{aligned}
\end{equation}
$\\$For any $\epsilon_1>0$, we can find small enough $\Delta \hat{t}$, such that for $\Delta t\leq\Delta \hat{t}$ $\sup\limits_{t\in \mathbb{R}}||\eta_T(t)-\hat{\eta}_t(t)||<\epsilon_1$, then
\begin{equation}
	\begin{aligned}
		||\eta(t)-\hat{\eta}_N(t)||&\leq ||\eta_N(t)-\hat{\eta}_N(t)||+||\eta(t)-\eta_N(t)||\\
		&\leq\beta e^{-\alpha N\Delta t}+\gamma\sup\limits_{t\in \mathbb{R}}|w(t)|+\epsilon_1
	\end{aligned}
\end{equation}
Take
\begin{equation}
	\begin{aligned}
		u(t)\triangleq&\frac{1}{k}[y_d^{(r)}(t)-\mathbf{r}^T\xi_d(t)-\mathbf{s}^T\eta(t)-g\cdot w(t)]\\
		\hat{u}(t)\triangleq&\frac{1}{k}[y_d^{(r)}(t)-\mathbf{r}^T\xi_d(t)-\mathbf{s}^T\hat{\eta}_N(t)]
	\end{aligned}
\end{equation}
Then
\begin{equation}
	\begin{aligned}
		|u(t)-\hat{u}(t)|<&\frac{1}{k}(||\mathbf{s}||(\beta e^{-\alpha N\Delta t}+\epsilon_1)+\\&(||\mathbf{s}||\gamma+g)\sup\limits_{t\in \mathbb{R}}|w(t)|)
	\end{aligned}
\end{equation}
For some initial state $\mathbf{x}(0)$, under inpute $u(t)$,we have, the outpute $y(t)$ can be strictly equal to $y_d(t)-h(t)$, then
\begin{equation}
	\begin{aligned}
		y_d(t)&=\mathbf{C}\mathbf{x}(t)\\&=\mathbf{C}(e^{\mathbf{A}t}\mathbf{x}(0)+\int_{0}^{t}e^{\mathbf{A}(t-\tau)}(\mathbf{B}u(\tau)+\mathbf{G}w(\tau))d\tau)
	\end{aligned}
\end{equation}
Assume the true system has initial states $\hat{\mathbf{x}}(0)$, under the outpute $\hat{u}(t)$, take
\begin{equation}
	\begin{aligned}
		\hat{y}_d(t)&=C\mathbf{x}(t)\\&=C(e^{At}\hat{\mathbf{x}}(0)+\int_{0}^{t}e^{A(t-\tau)}(B\hat{u}(\tau)+Gw(\tau))d\tau)
	\end{aligned}
\end{equation}
Then
\begin{align*}
	|\hat{y}_d(t)-y_d(t)|=&|\mathbf{C}(e^{\mathbf{A}t}(\hat{\mathbf{x}}(0)-\mathbf{x}(0))+\\&\int_{0}^{t}e^{\mathbf{A}(t-\tau)}(\hat{u}(\tau)-u(\tau))d\tau|\\
	\leq&||\mathbf{C}||(\beta_3e^{-t\alpha_3}\cdot||\mathbf{x}(0)-\hat{\mathbf{x}}(0)||+\\&\frac{1}{k}(||\mathbf{s}||(\beta e^{-\alpha N\Delta t}+\epsilon_1)+\\&(||s||\gamma+g)\sup\limits_{t\in \mathbb{R}}|w(t)|)\frac{\beta_3}{\alpha_3}(1-e^{-\alpha_3t}))
\end{align*}
The true output
\begin{equation}
	\begin{aligned}
		y(t)=\hat{y}_d(t)+h(t)
	\end{aligned}
\end{equation}
\begin{align*}
	&\lim\limits_{t\to+\infty}|y(t)-y_d(t)|\\\leq&\lim\limits_{t\to+\infty}|y(t)-\hat{y}_d(t)|+|\hat{y}_d(t)-y_d(t)|\\
	\leq&\sup\limits_{t\in \mathbb{R}}|h(t)|+||\mathbf{C}||\frac{||\mathbf{s}||\beta_3}{k\alpha_3}(\epsilon_1+\beta e^{-\alpha N\Delta t})+\\&\frac{\beta_3}{\alpha_3}(||\mathbf{s}||\gamma+g)\sup\limits_{t\in \mathbb{R}}|w(t)|
\end{align*}$\\$
Take $\epsilon_1$ small enough and $\hat{N}$ such that $||\mathbf{C}||\frac{||\mathbf{s}||\beta_3}{k\alpha_3}(\epsilon_1+\beta e^{-\alpha N\Delta t})<\epsilon$ and $\delta=\frac{\beta_3}{\alpha_3}(||\mathbf{s}||\gamma+g)$ $\blacksquare\\ \\$
$\textbf{B. Data-driven pseudoinverse approach for Koopman-}\\$
$\textbf{type operator parameter optimization design}$

	Theorem 1 illustrates that for the system with disturbance, if a suitable $\mathbf{K}$ can be found, then our Koopman-type operator can control this system with error depending on the disturbances. In this part we will show how to find $\mathbf{K}$ by data-driven pseudoinverse method.
Using every function $\phi_i$ in $\Phi$ as the input to the system and measuring the output $y_i$ of the system. Denote the output vector by $\mathbb{O}_i$
\begin{equation}
	\begin{aligned}
		&\mathbb{O}_i=[\phi_i(t_1),\phi_i(t_2)\cdots\phi_i(t_j)], \\&j\in\mathbb{Z}^+ and \ t_1<t_2\cdots<t_j
	\end{aligned}
\end{equation}
$\textit{Remark 2:}$
	For different input functions, $\phi_i$, $t_1,t_2,...,t_j$ are same and $t_1$ should be large enough to reduce the effect of the initial dynamic.$\\$
Take
\begin{equation}
	\mathbb{O}=[\mathbb{O}_1^T,\mathbb{O}_2^T,...]^T
\end{equation}
\begin{equation}
	\mathbb{O}_d=[y_d(t_1),y_d(t_2),...,y_d(t_j)]
\end{equation}
We calculate K by minimize
\begin{equation}
	||\mathbb{O}_d-\mathbf{K}^T\mathbb{O}||
\end{equation}
and the solution to this optimal problem can be written in exact form
\begin{equation}
	\mathbf{K}^T=\mathbb{O}_d\mathbb{O}^\dagger
\end{equation}
$\textbf{C. Improved Koopman-type method to handle large }\\$
$\textbf{disturbances}$

Monte Carlo methods tend to perform well in dealing with stochastic problems. For systems with large perturbations, the parameters of the Koopman-type operator obtained by the data-driven method may be very inaccurate, this section we will prove the errors in the data-driven process can be reduced using Monte Carlo methods.$\\$
$\\ \textbf{Theorem 2:}$
	Assume $E[\mathbb{O}]$ is the expection of the output matrix, and $\mathbb{O}^1,\mathbb{O}^2,...$ are the output matrix obtained from system (5), take $\overline{\mathbb{O}}_N=\frac{1}{N}\sum^N_{n=1}\mathbb{O}^n$. For given $\Phi$, if $E[r(t)]=E[w(t)]=0$, and $E[\int_0^tw^2(t)]<\infty$ for any $t\in\mathbb{R}$ then there exists $\sigma>0$, with probability $1-p$, $||\overline{\mathbb{O}}_N-E[\mathbb{O}]||_\infty< \frac{\sigma}{\sqrt{Np}}$.

$\\$
$\textbf{Proof}$ Denote $y_s$ as the actual output of system (1)
\begin{equation}
	\begin{aligned}
		&Var(y_s(t)-E[y_s(t)])\\&=Var(\int_{0}^{t}e^{\mathbf{A}(t-\tau)}\mathbf{G}w(\tau)d\tau)+Var(r(t))\\
		&=E[(\int_{0}^{t}e^{\mathbf{A}(t-\tau)}\mathbf{G}w(\tau)d\tau)^2]+Var(r(t))\\
		&\leq ||\mathbf{G}||^2(\int_{0}^{t}(\kappa_3e^{-\alpha_3(t-\tau)})^2d\tau) E[\int_{0}^{t}w^2(\tau)d\tau]+\\
		&Var(r(t))\\
		&\leq \frac{\kappa_3^2||\mathbf{G}||^2}{2\alpha_3}E[\int_{0}^{t}w^2(\tau)d\tau]+Var(r(t))
	\end{aligned}
\end{equation}
The second to the last inequality hold by Cauchy Inequality. Take
\begin{equation}
	V^2=\frac{\kappa_3^2||\mathbf{G}||^2}{2\alpha_3}\sup\limits_{t\in [0,t_j]}E[\int_{0}^{t}w^2(\tau)d\tau]+\sup\limits_{t\in \mathbb{R}}Var(r(t))
\end{equation}
Then $Var(y_s(t)-E[y(t)])\leq V^2$ for all $t\in [0,t_j]$, 
by Chebyshev's Inequality, with probablity $1-sp$, where s is the number of elements in matrix $\mathbb{O}$
\begin{equation}
	\begin{aligned}
		||\overline{\mathbb{O}}_N-E[\mathbb{O}]||_\infty&=||\frac{1}{N}\sum^N_{n=1}(\mathbb{O}^n-E[\mathbb{O}])||_\infty\\
		&\leq\frac{V}{\sqrt{Np}}
	\end{aligned}
\end{equation}
Take $\sigma=V\sqrt{s}$ $\blacksquare\\$
$\textit{Remark 3}$
	According to this theorem, for systems subjected to substantial disturbances, an effective strategy to mitigate the error and attain an accurate estimation of $\mathbf{K}$ is to employ averaging over multiple measurements.

\section{SIMULATION RESULTS}
In this section, we present a linear NMP system and validate the results in Section IV through simulation. In Part A, the system's parameters and the simulation methodology are designed. In Part B, data-driven approach in Section IV, Part B is applied to determine Koopman-type operator's parameters. Then the function provided by the Koopman operator is applied to the system to confirm the tracking accuracy demonstrated in Section IV, Part A. Lastly, in Part C, we corroborate that the Improved Koopman-type method proposed in Section IV, Part C is more effective in systems with substantial perturbations.$\\ \\$
$\textbf{A. Simulation system and parameter design}$

The simulation system we choose is
\begin{equation}
	\begin{aligned}
		\dot{\mathbf{x}}&=
		\begin{pmatrix}
			0 & 1 & 0 & 0\\
			0 & 0 & 1 & 0\\
			0 & 0 & 0 & 1\\
			-1 & -4 & -5.5 & -3.5
		\end{pmatrix}
		\mathbf{x}+
		\begin{pmatrix}
			0\\0\\0\\1
		\end{pmatrix}
		u(t)+\mathbf{G}w\\
		y&=
		\begin{pmatrix}
			-2 & 1 & 1 &0
		\end{pmatrix} x+h
	\end{aligned}
\end{equation}
Where $\mathbf{G}$ is a $1\times 4$ matrix. And $w,h$ are bounded disturbances, heir characteristics will be determined according to the specific simulation requirements. The system is non-minimum phase, and it has relative degree 2. The initial stste at $t=0$ is set to $\mathbf{x}(0)=\textbf{0}$.

The function being tracked is set to $y_d=\sin(0.1t)$. We take $T=10, \Delta t=0.5$ in Theorem 1. The output vector $\mathbb{O}_i$ are generated by $y_d(t+T-i\Delta t)=\sin(0.1(t+10-0.5i))$ as the input of the system for $i\in\{1,2\cdots 40\}$. And $\mathbb{O}_{41}$, $\mathbb{O}_{42}$ are generated by $\dot{y}_d=0.1\cos(0.1t)$, $\dot{y}_d=-0.01\sin(0.1t)$. All these output vector has $t_j=50+0.5j$ for $j\in{1,2\cdots100}$ in (19). Parameters of the Koopman-type operator are calculated by (25).

For a given input function $u(t)$, we calculate the value at y(t) by
\begin{equation}
	y(t)=\mathbf{C}(\int_0^te^{\mathbf{A}\tau}(\mathbf{B}u(\tau)+\mathbf{G}w(\tau)))+h(t)
\end{equation}
For the calculation of the integral we use partition method, in order to ensure the accuracy and do not occupy too much memory, the length of partition length is 0.01, which is accurate compare to $\Delta=0.05$.$\\ \\$
$\textbf{B. Performance of Koopman-type operator in}\\$
$\textbf{disturbanced system under data-driven pseudoinverse}$
$\textbf{approach}$

In this part, random disturbances are added to both the input and output. We take $\mathbf{G}$=(0 0 0 1)$^T$, $w(t)=U[0.05|u(t)|], h(t)=U[0.05|y(t)|]$. Solving the Koopman operator parameters by Data-driven method and data simulation are performed in the perturbed system. Figures 1 and 2 show the simulation results and errors for this disturbed system, and the images vary slightly from simulation to simulation due to the presence of random functions in the system. From Figure 2, the tracking error is less than $0.05max(y_d(t))=0.05$.
\begin{figure}[htbp]
	\includegraphics[width=8cm] {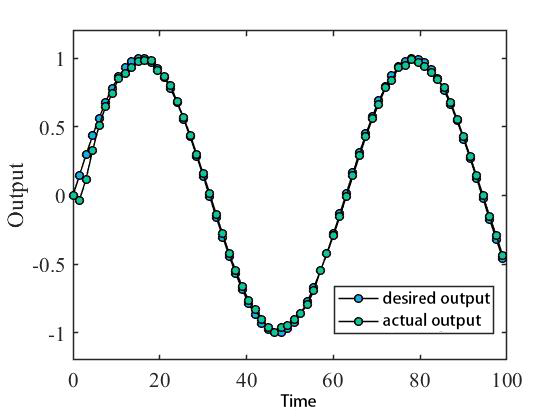}
	\caption{\label{1} Simulation result for disturbed system}
\end{figure}
\begin{figure}[H]
	\includegraphics[width=8cm] {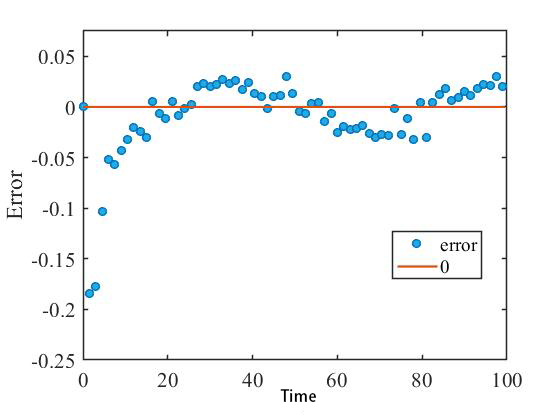}
	\caption{\label{1} Tracking error for undisturbed system}
\end{figure}$\\$
$\textbf{C. Performance of improved Koopman-type operator }\\$

In this part, we introduce a simulated scenario in which large disturbances are added to both the input and output of the system during the data-driven approach for solving the Koopman-type operator coefficients. Specifically, we set $\mathbf{G}$ to be (0 0 0 1)$^T$, and adopt $w(t)=U[0.2|u(t)|]$ and $h(t)=U[0.2|y(t)|]$ to model the larger errors in measurement that are often encountered in practical applications. To obtain different values of $K$, we select values of $N$ from 1, 5, and 10 in Theorem 2. After data-driven process, we get a 
parameter vector $\mathbf{K}$, then we input this K into the Koopman-type operator, and the resulting input function is used as the input to the unperturbed system, the error with the tracked function is obtained. Figure 3 illustrates the impact of varying Monte Carlo sampling times on the resulting tracking performance.$\\ \\$
\begin{figure}[H]
	\includegraphics[width=8cm] {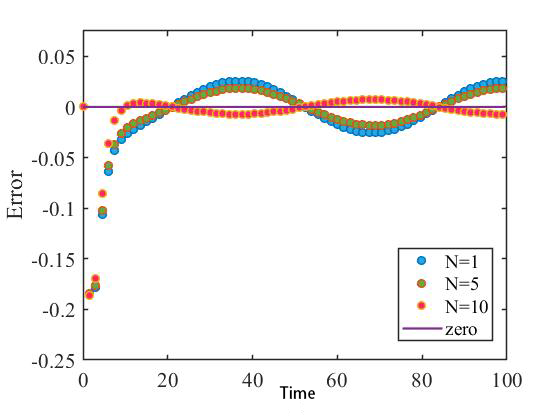}
	\caption{\label{1} The tracking error of Koopman-type parameter K under the disturbed system obtained by conducting different number of experiments in disturbed system}
\end{figure}
\section{CONCLUSIONS}
In this paper, we design a data-driven Koopman-type operator for linear non-minimum phase (NMP) systems, and give theoretical proof of the control accuracy of NMP systems using the Koopman-type method, this is the first proof for Koopman-type method in NMP systems. We demonstrate that the tracking accuracy of our Koopman-type operator is directly proportional to the ground-bound of the NMP system perturbation, given sufficiently long causal and non-causal desired outputs and appropriately small sampling intervals. Additionally, applying the Monte Carlo method helps mitigate the impact of perturbations on measurement results during the data-driven process. Our ongoing research efforts focus on extending this approach to address linear time-varying non-minimum phase systems.

\addtolength{\textheight}{-12cm}   
\bibliographystyle{plain}
\bibliography{ref}

\end{document}